\documentclass[showpacs,aps,graphicx]{revtex4}%,twocolumn
\usepackage{amsmath}%preprint,
\usepackage{graphicx}

\begin{document}

\title{Complete nondestructive analysis of two-photon six-qubit hyperentangled Bell states  assisted by cross-Kerr
nonlinearity\footnote{Published in Sci. Rep. \textbf{6}, 22016 (2016);}}

\author{Qian Liu,  Guan-Yu Wang,  Qing Ai,  Mei Zhang,  and Fu-Guo Deng\footnote{Corresponding author: fgdeng@bnu.edu.cn}}

\address{ Department of Physics, Applied Optics Beijing Area Major Laboratory,
Beijing Normal University, Beijing 100875, China}

\date{\today }

\begin{abstract}
Hyperentanglement, the entanglement in several degrees of freedom
(DOFs) of a quantum system, has attracted much attention as it can
be used to increase both the channel capacity of quantum
communication and its security largely. Here, we present the first
scheme to completely distinguish the hyperentangled Bell states of
two-photon systems in three DOFs with the help of cross-Kerr
nonlinearity without destruction, including two longitudinal
momentum DOFs and the polarization DOF. We use cross-Kerr
nonlinearity to construct quantum nondemolition detectors which can
be used to make a parity-check measurement and analyze Bell states
of two-photon systems in different DOFs.  Our complete scheme for
two-photon six-qubit hyperentangled Bell-state analysis may be
useful for the practical applications in quantum information,
especially in long-distance high-capacity quantum communication.
\end{abstract}
\pacs{03.67.Hk,  03.67.Bg, 42.50.Pq}

\maketitle

%03.67.Dd,

\section{Introduction}
\label{sec1}

Quantum entanglement plays an important role in quantum information
processing. It is the key resource for quantum communication tasks,
such as quantum teleportation \cite{teleportation}, quantum swapping
\cite{swapping}, quantum dense coding \cite{densecoding,super2},
quantum key distribution \cite{Ekert91,BBM92}, quantum secret
sharing \cite{QSS1}, quantum secure direct communication
\cite{Longliu,twostep,onetime,QSDCrev,longgfQSDC}, and so on.
Recently, hyperentanglement, the entanglement in multiple degrees of
freedom of a quantum system \cite{heper1,heper2,heper3}, has
attracted much attention. It can be used to complete the
deterministic entanglement purification for nonlocal photonic
systems in the polarization degree of freedom (DOF)
\cite{EPPsheng2,EPPsheng3,EPPlixh,EPPdeng,shengsr}, which reduces
largely the resource consumed for quantum repeaters. As it is
impossible to deterministically distinguish the four Bell states in
polarization with only linear optical elements, hyperentanglement
can also used to assist the complete Bell-state analysis (BSA)
\cite{EPPsheng2,EPPsheng3}. For instance, Kwiat and Weinfurter
\cite{Kwiat} proposed a BSA scheme using photons entangled in
polarization and momentum (spatial mode) in 1998. In 2003, Walborn
\emph{et al.} \cite{Walborn} presented a simple linear-optical
scheme for the complete Bell-state analysis of photons with
hyperentanglement in both polarization and momentum. The experiments
of a complete BSA with polarization-time-bin hyperentanglement
\cite{Schuck} and polarization-momentum hyperentanglement
\cite{Barbieri1} have also been reported in succession. For all the
linear-optical BSA protocols mentioned above, they use one DOF as an
ancillary to accomplish the complete BSA in the other DOF, rather
than distinguish all the hyperentangled Bell states themselves. In
2007, Wei \emph{et al}. \cite{HBSA} pointed out that 7 states in the
group of 16 orthogonal hyperentangled Bell states can be
distinguished with only linear optics. The general theoretical
explanation has been presented by Pisenti's group \cite{Pisenti} in
2011.

Hyperentanglement  of photon systems can  increase both the channel
capacity of long-distance quantum communication and its security. In
2008, Barreiro \emph{et al.} \cite{Barreiro1} beat the channel
capacity limit for linear photonic superdense coding with
polarization-orbital-angular-momentum hyperentanglement. In 2012,
Wang, Song, and Long \cite{repeater} proposed an efficient quantum
repeater protocol for long-distance quantum communication with
hyperentanglement. In 2013, Ren, Du, and Deng \cite{hyperECP} gave
the first hyperentanglement concentration protocol (hyper-ECP) for
two-photon four-qubit systems with linear optics. In the same year,
Ren and Deng \cite{HEPP1} proposed the original hyperentanglement
purification protocol (HEPP) for polarization-spatial hyperentangled
states assisted by diamond nitrogen-vacancy centers inside photonic
crystal cavities. In 2014, Ren, Du, and Deng \cite{twostepHEPP} gave
a two-step HEPP for polarization-spatial hyperentangled states with
the quantum-state-joining method, and it has a far higher
efficiency. Ren and Long \cite{hyperECPgeneral} proposed a general
hyper-ECP for photon systems assisted by quantum dot spins inside
optical microcavities.  Li and Ghose \cite{hyperECP1} presented a
hyper-ECP for multipartite entanglement via linear optics. Some
other interesting protocols for hyperentanglement concentration and
hyperentanglement purification \cite{hyperEPP2,hyperEPP3,hyperEPP4}
were presented in 2015.

In fact, in long-distance high-capacity quantum communication, the
complete analysis for all the orthogonal hyperentangled Bell states
of photon systems in multiple DOFs is necessary.  The 16 orthogonal
hyperentangled Bell states of two-photon systems in two DOFs can be
distinguished completely if nonlinear optics is introduced. In 2010,
Sheng \emph{et al.} \cite{kerr} gave the first scheme for the
complete hyperentangled-Bell-state analysis (HBSA) for quantum
communication with the help of cross-Kerr nonlinearity. In 2012, Ren
\emph{et al.} \cite{HBSA2} proposed another complete HBSA scheme for
photon systems in both the polarization and the spatial-mode DOFs
with the help of giant nonlinear optics in one-sided
quantum-dot-cavity systems. Using double-sided quantum-dot-cavity
systems, the complete HBSA scheme also can be accomplished
\cite{HBSA3}.  Xia \emph{et al.} \cite{Xia} proposed an efficient
scheme for hyperentangled Greenberger-Horne-Zeilinger-state analysis
with cross-Kerr nonlinearity.  Recently, the hyperentangled Bell
states for two-photon six-qubit systems were produced in experiments
\cite{Ceccarelli,Vallone}, but there are no schemes for the complete
analysis on two-photon six-qubit quantum states as they are far more
difficult, compared with the Bell states in both one and two DOFs.

In this paper, we give the first scheme to completely distinguish
the hyperentangled Bell states of two-photon systems in three DOFs
with the help of cross-Kerr nonlinearity without destruction,
including a polarization DOF and  double longitudinal momentum DOFs.
Our HBSA protocol for two-photon six-qubit hyperentangled systems
may be useful in the practical applications in quantum information
processing, blind quantum computation, distributed quantum
computation, and especially long-distance high-capacity quantum
communication in the future. With hyperdense coding on two-photon
systems entangled in three DOFs simultaneously as an example, we
show the principle of the applications of our HBSA protocol in
detail.

\begin{figure}[!h]%[tpb]  % picture 1
\begin{center}
\includegraphics[width=12 cm,angle=0]{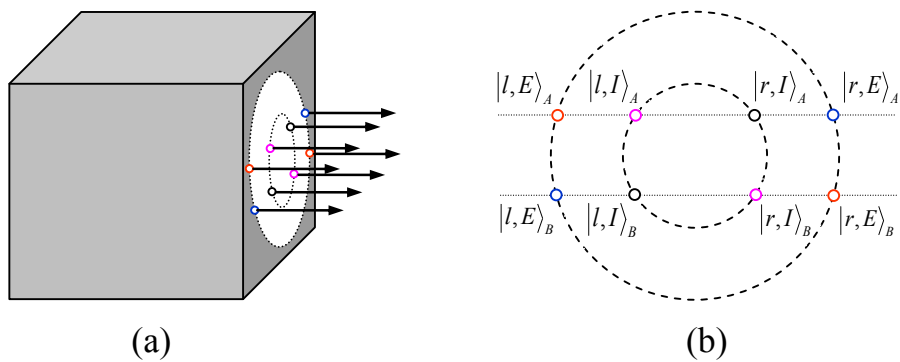}
\caption{ (a) Source for  two-photon  six-qubit hyperentangled Bell
states. A detailed description of the source is given in the
previous work \cite{Vallone}. (b) Modes for two-photon six-qubit
hyperentangled Bell states. The upper modes correspond to Alice's
photon, while the lower modes correspond to Bob's photon. $l$, $r$,
$I$, and $E$ are the left, right, internal, and external modes for a
photon, respectively.}\label{fig1}
\end{center}
\end{figure}

\section{Complete  analysis scheme for two-photon six-qubit hyperentangled Bell states} \label{sec3}

\subsection{Complete  analysis  for the states of a two-photon system in momentum modes}
\label{sec31}

A hyperentangled Bell state of two-photon six-qubit systems in three
DOFs can be described as follows:
\begin{eqnarray}\label{eq1}                    %  Eq 14
\begin{split}
 |HE_{6}\rangle=&\frac{1}{\sqrt{2}}(|H\rangle_{A}|H\rangle_{B}+|V\rangle_{A}|V\rangle_{B})
 \otimes\frac{1}{\sqrt{2}} (|l\rangle_{A}|r\rangle_{B}+|r\rangle_{A}|l\rangle_{B}) \\
 &\otimes\frac{1}{\sqrt{2}} (|I\rangle_{A}|I\rangle_{B}+|E\rangle_{A}|E\rangle_{B}).
 \end{split}
\end{eqnarray}
Here the subscripts A and B denote the two photons. $H$ and $V$
represent the horizontal and the vertical polarizations of photons,
respectively. The three independent DOFs are polarization and a
double longitudinal momentum ($r/l$ and $E/I$), shown in Fig. 1. The
system of the two-photon six-qubit source \cite{Vallone} consists of
two type-$\uppercase\expandafter{\romannumeral 1}$ $\beta$ barium
borate (BBO) crystal slabs and a eight-hole screen. When a
continuous-wave (cw) vertically polarized $Ar^{+}$ laser beam
interacts through spontaneous parametric down-conversation (SPDC)
with the two BBO crystal slabs, and the nonlinear interaction
between the laser beam and the BBO crystal leads to the production
of the degenerate photon pairs, which are entangled in polarization
and belong to the surfaces of two emission cones. As shown  in Fig.
1 (a), the insertion of a eight-hole screen allows us to achieve the
double longitudinal momentum entanglement. The labels in Fig. 1 (b)
are used to identify the selected modes. The internal $(I)$ and the
external $(E)$ cones correspond  to the first and the second
crystals, respectively. Furthermore, $l$ ($r$) refers to the left
(right) side of each cone.

The distinction between the internal ($I$) and the external ($E$)
modes provides us the second longitudinal momentum DOF, while the
first longitudinal momentum DOF comes from the distinction between
the left ($l$) and right ($r$) modes. Therefore, the six-qubit
hyperentangled state described in Eq.(\ref{eq1})  is given by the
product of one polarization entangled state and two longitudinal
momentum entangled states of a photon pair.

Let us denote the four Bell states in the polarization DOF of
two-photon systems as
\begin{eqnarray}  %  Eq 1
\begin{split}
|\phi^{\pm}\rangle_{P}&=\frac{1}{\sqrt{2}}(|H\rangle_{A}|H\rangle_{B}\pm|V\rangle_{A}|V\rangle_{B}),\\
|\psi^{\pm}\rangle_{P}&=\frac{1}{\sqrt{2}}(|H\rangle_{A}|V\rangle_{B}\pm|V\rangle_{A}|H\rangle_{B}),
\end{split}
\end{eqnarray}
and four Bell states in the first longitudinal momentum DOF as
\begin{eqnarray}  %  Eq 2
\begin{split}
|\phi^{\pm}\rangle_{F}&=\frac{1}{\sqrt{2}}(|l\rangle_{A}|l\rangle_{B}\pm|r\rangle_{A}|r\rangle_{B}),\\
|\psi^{\pm}\rangle_{F}&=\frac{1}{\sqrt{2}}(|l\rangle_{A}|r\rangle_{B}\pm|r\rangle_{A}|l\rangle_{B}),
\end{split}
\end{eqnarray}
while the four Bell states in the second longitudinal momentum DOF
can be expressed as
\begin{eqnarray}  %  Eq 3
\begin{split}
|\phi^{\pm}\rangle_{S}&=\frac{1}{\sqrt{2}}(|I\rangle_{A}|I\rangle_{B}\pm|E\rangle_{A}|E\rangle_{B}),\\
|\psi^{\pm}\rangle_{S}&=\frac{1}{\sqrt{2}}(|I\rangle_{A}|E\rangle_{B}\pm|E\rangle_{A}|I\rangle_{B}).
\end{split}
\end{eqnarray}
Here the subscripts  $P$, $F$, and $S$ denote the polarization,  the
first  longitudinal momentum, and the second longitudinal momentum
DOFs of a two-photon six-qubit system, respectively.

The principle of our scheme for the complete analysis on the quantum
states of a two-photon six-qubit system in the first longitudinal
momentum DOF is shown in Fig. 2. In detail, one can let the two
photons $AB$ pass through the first quantum nondemolition detector
(QND$_1$) whose circuit is shown in Fig. 2 (a). Based on the
principle of cross-Kerr effect (see Methods), the evolution of
two-photon six-qubit hyperentangled Bell states and the coherent
state can be described as follows:
\begin{eqnarray}   %  Eq 4
\begin{split}
|P\rangle|\phi^{\pm}\rangle_{F}|S\rangle|\alpha\rangle&=\frac{1}{\sqrt{2}}|P\rangle|S\rangle
\left(|l\rangle_{A}|l\rangle_{B}\pm|r\rangle_{A}|r\rangle_{B}\right)|\alpha\rangle\\
&\rightarrow
\frac{1}{\sqrt{2}}|P\rangle|S\rangle\left(|l\rangle_{A}|l\rangle_{B})|\alpha
e^{-2i\theta}\rangle\pm|r\rangle_{A}|r\rangle_{B}|\alpha
e^{2i\theta}\rangle\right),\\
|P\rangle|\psi^{\pm}\rangle_{F}|S\rangle|\alpha\rangle&=\frac{1}{\sqrt{2}}|P\rangle|S\rangle
(|l\rangle_{A}|r\rangle_{B}\pm|r\rangle_{A}|l\rangle_{B})|\alpha\rangle\\
&\rightarrow\frac{1}{\sqrt{2}}|P\rangle|S\rangle\left(|l\rangle_{A}|r\rangle_{B}|\alpha\rangle\pm|r\rangle_{A}|l\rangle_{B}|\alpha\rangle\right)\\
&=\frac{1}{\sqrt{2}}|P\rangle|S\rangle\left(|l\rangle_{A}|r\rangle_{B}\pm|r\rangle_{A}|l\rangle_{B}\right)|\alpha\rangle.\\
\end{split}
\end{eqnarray}
Here, $|P\rangle$ represents the four Bell stats in the polariztion
DOF, and $|S\rangle$ denotes the four Bell states in the second
longitudinal momentum DOF. The equation above shows that the Bell
states of other two DOFs have not changed. If these two photons are
in the same state $|r\rangle_{A}|r\rangle_{B}$ or
$|l\rangle_{A}|l\rangle_{B}$ in the first longitudinal momentum DOF,
the coherent probe beam will pick up a phase shift $+2\theta$ or
$-2\theta$. If these two photons are in the different  states
$|l\rangle_{A}|r\rangle_{B}$ or $|r\rangle_{A}|l\rangle_{B}$, the
phase shift of the coherent probe beam will be 0. As the homodyne
measurement cannot distinguish $+2\theta$ from $-2\theta$, there are
only two measurement outcomes $|\alpha\rangle$ and $|\alpha
e^{\pm2i\theta}\rangle$ for the coherent probe beam. Thus, according
to the measurement results, one can distinguish the even-parity
states $|\phi^{\pm}\rangle_{F}$ from the odd-parity states
$|\psi^{\pm}\rangle_{F}$. That is, QND$_1$ shown in Fig. 2 (a) is a
quantum nondemolition detector, with which one can distinguish the
parity of the two photons $A$ and $B$ in the first longitudinal
momentum DOF.

After QND$_1$, one can divide the four Bell states in the first
longitudinal momentum DOF into two groups, $|\phi^{\pm}\rangle_{F}$
and $|\psi^{\pm}\rangle_{F}$. The next task is to distinguish  the
different phases in $|\phi^{\pm}\rangle_{F}$ and
$|\psi^{\pm}\rangle_{F}$, respectively. By using the 50:50 beam
splitters (BSs) shown in Fig. 2 (b) on the photons, one can get the
following transformations:
\begin{eqnarray}   %  Eq 5
\begin{split}
|P\rangle|\phi^{+}\rangle_{F}|S\rangle&\;\;\rightarrow\;\;\frac{1}{\sqrt{2}}|P\rangle|S\rangle(|l\rangle_{A}|l\rangle_{B}+|r\rangle_{A}|r\rangle_{B}),\\
|P\rangle|\phi^{-}\rangle_{F}|S\rangle&\;\;\rightarrow\;\;\frac{1}{\sqrt{2}}|P\rangle|S\rangle(|l\rangle_{A}|r\rangle_{B}+|r\rangle_{A}|l\rangle_{B}),\\
|P\rangle|\psi^{+}\rangle_{F}|S\rangle&\;\;\rightarrow\;\;\frac{1}{\sqrt{2}}|P\rangle|S\rangle(|l\rangle_{A}|l\rangle_{B}-|r\rangle_{A}|r\rangle_{B}),\\
|P\rangle|\psi^{-}\rangle_{F}|S\rangle&\;\;\rightarrow\;\;\frac{1}{\sqrt{2}}|P\rangle|S\rangle(|l\rangle_{A}|r\rangle_{B}-|r\rangle_{A}|l\rangle_{B}).
\end{split}
\end{eqnarray}
As the BSs transform the phase difference of the two states from
each group into the parity difference, the two Bell states in the
same group will belong to different groups after the BSs. Then, if
we let photon A and photon B pass  through  the same quantum circuit
as QND$_1$ shown in Fig. 2 (b), the four Bell states can be
distinguished completely. Although the states
$|P\rangle|\phi^{-}\rangle_{F}|S\rangle$ and
$|P\rangle|\psi^{+}\rangle_{F}|S\rangle$ have changed into
$\frac{1}{\sqrt{2}}|P\rangle|S\rangle(|l\rangle_{A}|r\rangle_{B}+|r\rangle_{A}|l\rangle_{B})$
and
$\frac{1}{\sqrt{2}}|P\rangle|S\rangle(|l\rangle_{A}|l\rangle_{B}-|r\rangle_{A}|r\rangle_{B})$
by BSs in this procedure,  respectively, one can use other BSs after
the quantum circuit as QND$_{1}$ to recover the initial Bell states
in the first longitudinal momentum DOF. The relationship between the
measurement results of these two QNDs and the corresponding Bell
states in the first longitudinal momentum DOF is shown in Table
\ref{table1}.

Now, we have finished the distinction of the  four Bell states in
the first longitudinal momentum, without destroying the
hyperentanglement in the other two DOFs. Then we move to the next
step to distinguish the four Bell states in the second longitudinal
momentum DOF. As the first longitudinal momentum and the second
longitudinal momentum are all linear momentum, what we do to realize
the next distinction is similar to the analysis protocol of the
first longitudinal momentum DOF. The difference is to interchange
the path labels $r/l$ to $E/I$. The principle for distinguishing the
four Bell states of the two-photon system in the second longitudinal
momentum  DOF is shown in Fig. 3. Here, we let the two photons pass
through  QND$_3$ and then QND$_4$ in sequence. With these two QNDs,
we can analyze the four Bell states in the second longitudinal
momentum DOF completely. The relationship between the measurement
results of this scheme and the corresponding Bell states in the
second longitudinal momentum DOF are described in Table
\ref{table2}.

\begin{figure}%[!h]%[tpb]  % picture 1
\begin{center}
\includegraphics[width=10 cm,angle=0]{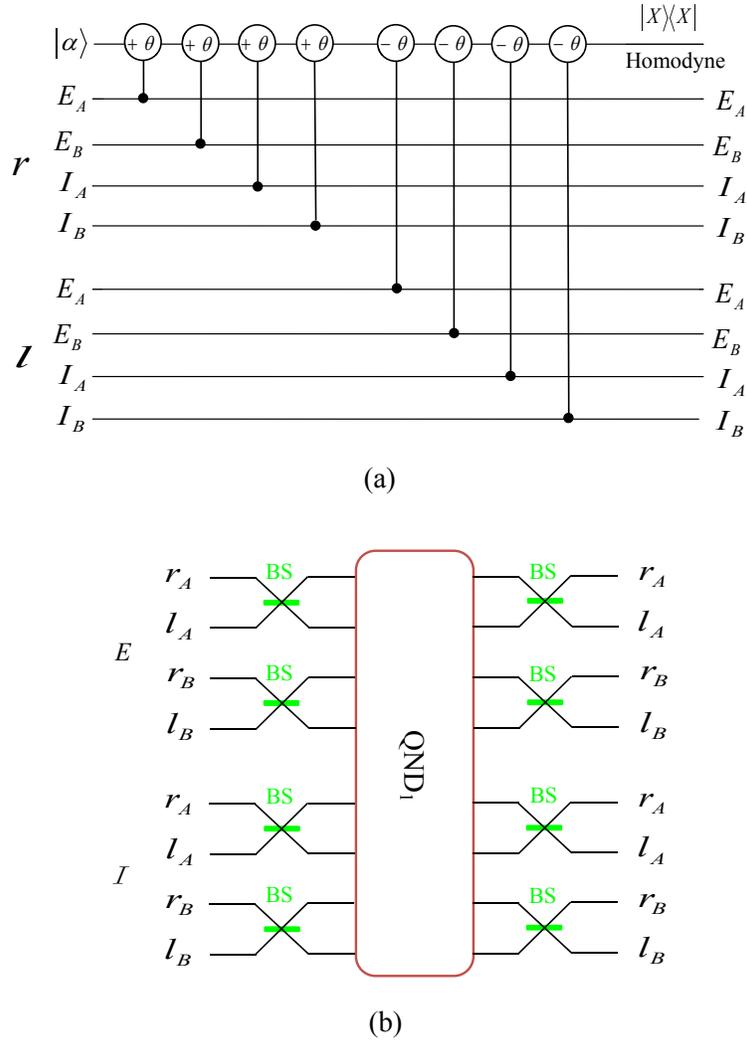}
\caption{ (a) Schematic diagram of QND$_1$ which is used to
distinguish the even-parity states $|\phi^{\pm}\rangle_{F}$ from the
odd-parity states $|\psi^{\pm}\rangle_{F}$ in the first longitudinal
momentum DOF of the two-photon six-qubit system $AB$. $\pm\theta$
denotes the cross-Kerr nonlinear media which will make the cohere
probe beam $\vert \alpha\rangle$ have a phase shift $\pm\theta$ when
there is a signal photon passing through it. $|X\rangle\langle X|$
is the homodyne measurement to discriminate different phase shifts
of the coherent probe beam. $r$ and $l$ represent the left and the
right sides of each cone from where the photons emit, respectively.
The internal $(I)$ and the external $(E)$ cones correspond  to the
first and the second crystal from which the photons are produced,
respectively. (b) Schematic diagram of QND$_2$.  Each of the 50:50
BSs acts as a Hadamard operation
($|r\rangle\rightarrow\frac{1}{\sqrt{2}}(|r\rangle)+|l\rangle$,
$|l\rangle\rightarrow\frac{1}{\sqrt{2}}(|r\rangle)-|l\rangle$) on
the photon in the first longitudinal momentum DOF. After these two
photons pass through the BSs,  one can use mirrors to separate the
paths of photons.}\label{fig2}
\end{center}
\end{figure}

\begin{figure}[tpb]  % picture 2
\begin{center}
\includegraphics[width=8.5 cm,angle=0]{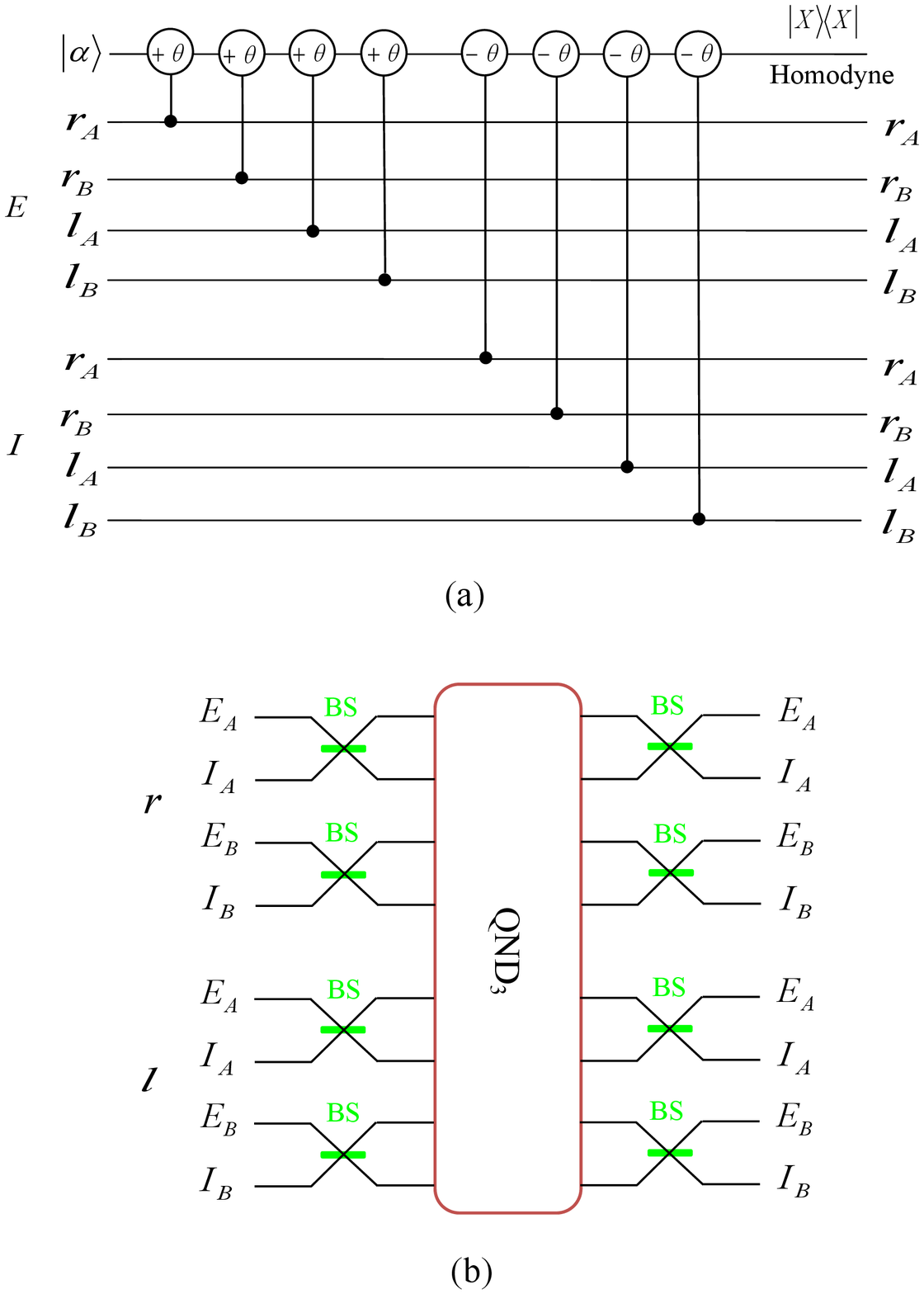}
\caption{  Schematic diagram for distinguishing the four Bell states
of the two-photon six-qubit system in the second longitudinal
momentum DOF. (a) QND$_3$. (b) QND$_4$, the 50:50 BS is used to
perform the Hadamard operation
($|E\rangle\rightarrow\frac{1}{\sqrt{2}}(|E\rangle)+|I\rangle$,
$|I\rangle\rightarrow\frac{1}{\sqrt{2}}(|E\rangle)-|I\rangle$) on
the second longitudinal momentum DOF of photons. }\label{fig3}
\end{center}
\end{figure}

The principle of our scheme for the complete  analysis  on  the
quantum states of a two-photon six-qubit system in the first
longitudinal momentum DOF is shown in Figs .\ref{fig2} and
\ref{fig3}. In detail, one can let the two photons $AB$ pass through
the first quantum nondemolition detector (QND$_1$) whose circuit is
shown in Fig.\ref{fig2}. Based on the principle of cross-Kerr
effects, the evolution  of the  four Bell states in the first
longitudinal momentum and the coherent state can be described as
follows:
\begin{eqnarray} %  Eq. 7
%\begin{split}
|\phi^{\pm}\rangle_{F}|\alpha\rangle& =&\frac{1}{\sqrt{2}}\left(|l\rangle_{A}|l\rangle_{B}\pm|r\rangle_{A}|r\rangle_{B}\right)|\alpha\rangle%\nonumber\\
\;\;\rightarrow\;\;
\frac{1}{\sqrt{2}}\left(|l\rangle_{A}|l\rangle_{B})|\alpha
e^{-2i\theta}\rangle\pm|r\rangle_{A}|r\rangle_{B}|\alpha
e^{2i\theta}\rangle\right),\;\;\;\;\;\;\;\;\;\;\;\;\;\;\;\;\;\;\;\;\;\;\;\;\;\;\;\;\;\;\;\;\;\;\;\;   %\\&\;\;\doteq
%\frac{1}{\sqrt{2}}\left(|l\rangle_{A}|l\rangle_{B})
%\pm|r\rangle_{A}|r\rangle_{B}\right)|\alpha
%e^{2i\theta}\rangle,%\nonumber\\
%\end{split}
\end{eqnarray}
\begin{eqnarray} %  Eq. 8
%\begin{split}
|\psi^{\pm}\rangle_{F}|\alpha\rangle=\frac{1}{\sqrt{2}}(|l\rangle_{A}|r\rangle_{B}\pm|r\rangle_{A}|l\rangle_{B})|\alpha\rangle
 \;\;\rightarrow\;\;
\frac{1}{\sqrt{2}}\left(|l\rangle_{A}|r\rangle_{B}|\alpha\rangle\pm|r\rangle_{A}|l\rangle_{B}|\alpha\rangle\right)
=\frac{1}{\sqrt{2}}\left(|l\rangle_{A}|r\rangle_{B}\pm|r\rangle_{A}|l\rangle_{B}\right)|\alpha\rangle.
%\end{split}
\end{eqnarray}
If these two photons are in the same state
$|r\rangle_{A}|r\rangle_{B}$ or $|l\rangle_{A}|l\rangle_{B}$ in the
first longitudinal momentum DOF, the coherent probe beam will pick
up a phase shift $+2\theta$ or $-2\theta$. If these two photons are
in the different  states $|l\rangle_{A}|r\rangle_{B}$ or
$|r\rangle_{A}|l\rangle_{B}$, the phase shift of the coherent probe
beam will be 0. As the homodyne measurement cannot distinguish
$+2\theta$ from $-2\theta$, there are only two measurement outcomes
$|\alpha\rangle$ and $|\alpha e^{\pm2i\theta}\rangle$ for the
coherent probe beam. Thus, according to the measurement results, one
can distinguish the even-parity states $|\phi^{\pm}\rangle_{F}$ from
the odd-parity states $|\psi^{\pm}\rangle_{F}$. That is,  QND$_1$
shown in Fig. \ref{fig2} is a quantum nondemolition detector, with
which one can distinguish the parity of the two photons $A$ and $B$
in the first longitudinal momentum DOF.

After QND$_1$, one can divided the four Bell states in the first
longitudinal momentum DOF into two groups, $|\phi^{\pm}\rangle_{F}$
and $|\psi^{\pm}\rangle_{F}$. The next task is to distinguish  the
different phases in $|\phi^{\pm}\rangle_{F}$ and
$|\psi^{\pm}\rangle_{F}$, respectively. By using the 50:50 beam
splitters (BSs) shown in Fig.\ref{fig3} on the photons, one can get
the following transformations:
\begin{eqnarray}  %  Eq. 9
%\begin{split}
|\phi^{+}\rangle_{F}&\;\;\rightarrow\;\;\frac{1}{\sqrt{2}}(|l\rangle_{A}|l\rangle_{B}+|r\rangle_{A}|r\rangle_{B}), \;\;\;\;\;\;\;\;\;\;%\\
|\phi^{-}\rangle_{F} \;\;\rightarrow\;\;\frac{1}{\sqrt{2}}(|l\rangle_{A}|r\rangle_{B}+|r\rangle_{A}|l\rangle_{B}), \nonumber\\
|\psi^{+}\rangle_{F}&\;\;\rightarrow\;\;\frac{1}{\sqrt{2}}(|l\rangle_{A}|l\rangle_{B}-|r\rangle_{A}|r\rangle_{B}), \;\;\;\;\;\;\;\;\;\;%\\
|\psi^{-}\rangle_{F}\;\;\rightarrow\;\;\frac{1}{\sqrt{2}}(|l\rangle_{A}|r\rangle_{B}-|r\rangle_{A}|l\rangle_{B}).
%\end{split}
\end{eqnarray}
One  can see that the BSs  transform the phase difference of the two
states from each group into the parity difference. The two Bell
states in the same group will belong to different groups after the
BSs. When these two photons pass  through the same quantum circuit
as QND$_1$ (that is, QND$_2$ shown in Fig. \ref{fig3}), the four
Bell states can be distinguished completely. The relationship
between the measurement results of these two QNDs and the
corresponding Bell states in the first longitudinal momentum DOF is
shown in Table \ref{table1}.

\begin{table}%[htb]
\centering \caption{The relationship between the four Bell states in
the first longitudinal momentum DOF and the measurement results of
QND$_1$ and QND$_2$.}
%\begin{ruledtabular}
\begin{tabular}{ccc}%{lp{1.2in} lp{1.2in}}
%\cline{1-3}
\hline\hline
   Bell states                  &     $\;\;\;\;\;\;\;\;$ QND$_1$ $\;\;\;\;\;\;\;\;$     &   QND$_2$      \\
   \hline
$|\phi^{+}\rangle_{F}$   &       $\pm2\theta$                      &   $\pm2\theta$\\
$|\phi^{-}\rangle_{F}$   &       $\pm2\theta$                      &   $0$ \\
$|\psi^{+}\rangle_{F}$   &       $0$                      &   $\pm2\theta$ \\
$|\psi^{-}\rangle_{F}$   &       $0$                      &   $0$\\
\hline\hline
\end{tabular}\label{table1}
%\end{ruledtabular}
\end{table}

\begin{figure}[tpb]  % picture 3
\begin{center}
\includegraphics[width=9.0cm,angle=0]{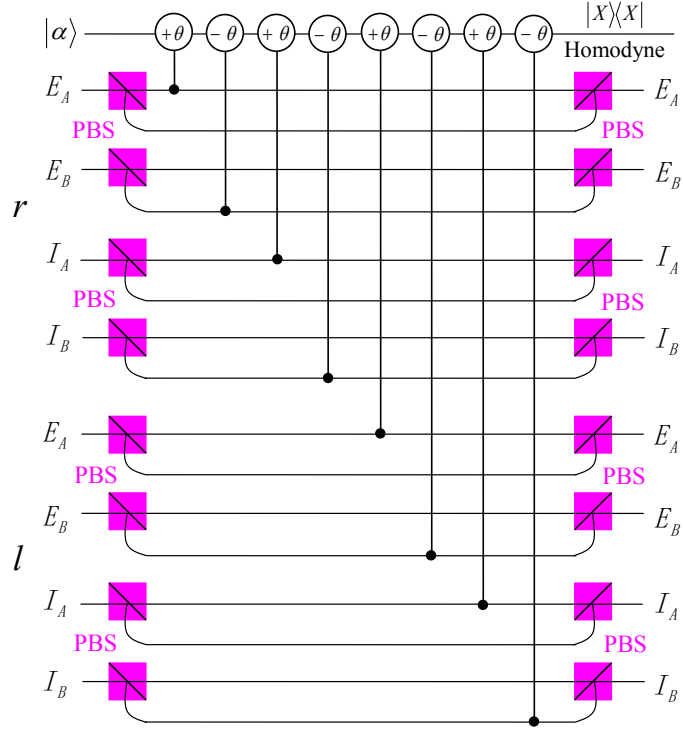}
\caption{Schematic diagram of QND$_5$ which is used to distinguish
the even-parity states $|\phi^{\pm}\rangle_{P}$ from the odd-parity
states $|\psi^{\pm}\rangle_{P}$ in polarization DOF of the
two-photon six-qubit system $AB$. PBS represents a polarizing beam
splitter which is used to transmit the horizontal ($H$) polarization
photon and reflect the vertical ($V$) polarization photon,
respectively.}\label{fig4}
\end{center}
\end{figure}

\begin{table}%[htb]
\centering \caption{The relationship between the four Bell states in
the second longitudinal momentum DOF and the measurement results of
QND$_3$ and QND$_4$.}
%\begin{ruledtabular}
\begin{tabular}{ccc}%{lp{1.2in} lp{1.2in}}
%\cline{1-3}
\hline\hline
   Bell states                  &     $\;\;\;\;\;\;\;\;$ QND$_3$ $\;\;\;\;\;\;\;\;$     &   QND$_4$      \\
   \hline
$|\phi^{+}\rangle_{S}$   &       $\pm2\theta$                      &   $\pm2\theta$\\
$|\phi^{-}\rangle_{S}$   &       $\pm2\theta$                      &   $0$ \\
$|\psi^{+}\rangle_{S}$   &       $0$                      &   $\pm2\theta$ \\
$|\psi^{-}\rangle_{S}$   &       $0$                      &   $0$\\
\hline\hline
\end{tabular}\label{table2}
%\end{ruledtabular}
\end{table}

\subsection{Complete six-qubit hyperentangled Bell state analysis scheme
for states in polarization} \label{sec32}

Now, let us move our attention to the last task, which is to
distinguish the four Bell states of the two-photon six-qubit system
in the polarization DOF. The analysis of the four Bell states in
polarization is analogous to that in previous works
\cite{Barrett,kerr}. The schematic diagram for the distinction of
the four Bell states in polarization is shown in Fig. 4 and Fig. 5.

According to QND$_{5}$ shown in Fig. 4, the states
$|\phi^{\pm}\rangle_{P}|F\rangle|S\rangle$ with the coherent state
$|\alpha\rangle$ evolve as
\begin{eqnarray}   %  Eq 6
\begin{split}
|\phi^{\pm}\rangle_{P}|F\rangle|S\rangle|\alpha\rangle
&=\frac{1}{\sqrt{2}}|F\rangle|S\rangle(|H\rangle_{A}|H\rangle_{B}\pm|V\rangle_{A}|V\rangle_{B})|\alpha\rangle\\
&\rightarrow\frac{1}{\sqrt{2}}|F\rangle|S\rangle\left(|H\rangle_{A}|H\rangle_{B}|\alpha
e^{i\theta}\rangle\pm|V\rangle_{A}|V\rangle_{B}|\alpha
e^{-i\theta}\rangle\right),
\end{split}
\end{eqnarray}
while the states $|\psi^{\pm}\rangle_{P}|F\rangle|S\rangle$ with the
coherent state $|\alpha\rangle$ evolve as
\begin{eqnarray}    %  Eq 7
\begin{split}
|\psi^{\pm}\rangle_{P}|F\rangle|S\rangle|\alpha\rangle
&=\frac{1}{\sqrt{2}}|F\rangle|S\rangle(|H\rangle_{A}|V\rangle_{B}\pm|V\rangle_{A}|H\rangle_{B})|\alpha\rangle\\
&\rightarrow\frac{1}{\sqrt{2}}|F\rangle|S\rangle(|H\rangle_{A}|V\rangle_{B}|\alpha\rangle\pm|V\rangle_{A}|H\rangle_{B}|\alpha\rangle)\\
&=\frac{1}{\sqrt{2}}|F\rangle|S\rangle(|H\rangle_{A}|V\rangle_{B}\pm|V\rangle_{A}|H\rangle_{B})|\alpha\rangle,\;\;\;\;
\end{split}
\end{eqnarray} where $|F\rangle$ represents the four Bell states in
the first longitudinal momentum DOF. In these evolutions, the modes
$|H\rangle_{A}|H\rangle_{B}$ or $|V\rangle_{A}|V\rangle_{B}$ will
let the coherent probe beam pick up a phase shift $+\theta$ or
$-\theta$, but the coherent probe beam will pick up no phase shift
if the two photons are in the mode $|H\rangle_{A}|V\rangle_{B}$ or
$|V\rangle_{A}|H\rangle_{B}$. With an X-quadrature measurement on
the coherent beam, as  $|\alpha e^{\pm i\theta}\rangle$ cannot be
distinguished, one can divide the four Bell states in polarization
into two groups, the even-parity one $\{\vert \phi^+\rangle_P, \vert
\phi^-\rangle_P\}$ and the odd-parity one $\{\vert \psi^+\rangle_P,
\vert \psi^-\rangle_P\}$.

The next step is to distinguish the different relative phases in
each of these two groups. This task can be accomplished with the
circuit shown in Fig. 5. Here the wave plate $R_{45}$ is used to
accomplish a Hadamard operation on the polarization of photons. A
Hadamard operation on each of the two photons $AB$ will make the
following transformations:
\begin{eqnarray}   %  Eq 8
\begin{split}
|\phi^{+}\rangle_{P}|F\rangle|S\rangle&\;\;\rightarrow\;\;\frac{1}{\sqrt{2}}|F\rangle|S\rangle\left(|H\rangle_{A}|H\rangle_{B}+|V\rangle_{A}|V\rangle_{B}\right),\\
|\phi^{-}\rangle_{P}|F\rangle|S\rangle&\;\;\rightarrow\;\;\frac{1}{\sqrt{2}}|F\rangle|S\rangle\left(|H\rangle_{A}|V\rangle_{B}+|V\rangle_{A}|H\rangle_{B}\right),\\
|\psi^{+}\rangle_{P}|F\rangle|S\rangle&\;\;\rightarrow\;\;\frac{1}{\sqrt{2}}|F\rangle|S\rangle\left(|H\rangle_{A}|H\rangle_{B}-|V\rangle_{A}|V\rangle_{B}\right),\\
|\psi^{-}\rangle_{P}|F\rangle|S\rangle&\;\;\rightarrow\;\;\frac{1}{\sqrt{2}}|F\rangle|S\rangle\left(|H\rangle_{A}|V\rangle_{B}-|V\rangle_{A}|H\rangle_{B}\right).
\end{split}
\end{eqnarray}
As $R_{45}$ can  transform the phase difference into the parity
difference, one can then use the same quantum circuit as  QND$_5$ to
distinguish the parity difference between the two states in each
group. Then we use other $R_{45}$ to recover the initial Bell states
in polarization DOF. That is, after the photons pass through QND$_6$
shown in Fig. 5, the two Bell states in the even-parity group
$\{\vert \phi^+\rangle_P, \vert \phi^-\rangle_P\}$ or the odd-parity
one $\{\vert \psi^+\rangle_P, \vert \psi^-\rangle_P\}$ can be
distinguished completely. The relationship between the measurement
results of this scheme and the corresponding Bell states in
polarization is described in Table \ref{table3}.

From the analysis above, one can see that the complete
nondestructive analysis for two-photon six-qubit hyperentangled Bell
states can be accomplished with the sequential connection of the six
QNDs. This complete HBSA can be used to complete some other
important tasks in high-capacity quantum communication, such as
teleportation with photon systems in three DOFs, hyperentanglement
swapping, quantum hyperdense coding, and so on.

\begin{figure}[!h]%[tpb]  % picture 4
\begin{center}
\includegraphics[width=7cm,angle=0]{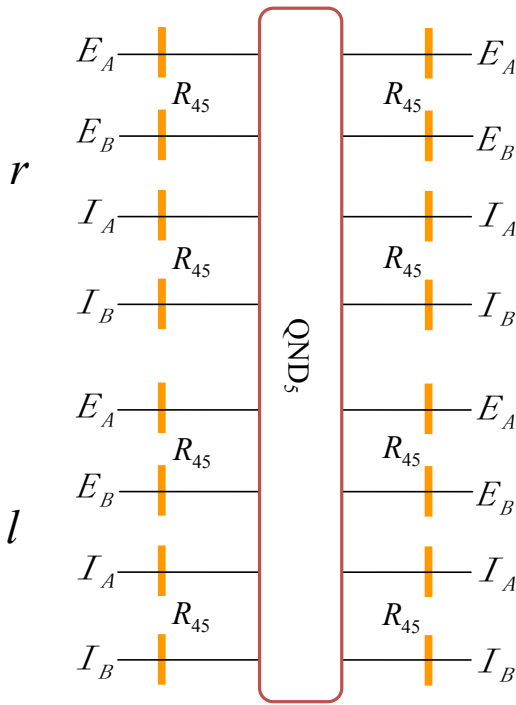}
\caption{Schematic diagram of QND$_6$. $R_{45}$ represents the wave
plate which rotates the horizontal and vertical polarizations by
$45^{o}$ to accomplish a Hadamard operation
($|H\rangle\rightarrow\frac{1}{\sqrt{2}}(|H\rangle)+|V\rangle$,
$|V\rangle\rightarrow\frac{1}{\sqrt{2}}(|H\rangle)-|V\rangle$) on
polarization of photons. }\label{fig5}
\end{center}
\end{figure}

\begin{table} %[!htbp]
\centering \caption{The relationship between the four Bell states in
the polarization DOF and the measurement results of QND$_5$ and
QND$_6$.}
%\begin{ruledtabular}
\begin{tabular}{ccc}%{lp{1.2in} lp{1.2in}}
%\cline{1-3}
\hline\hline
   Bell states                  &     $\;\;\;\;\;\;\;\;$ QND$_5$ $\;\;\;\;\;\;\;\;$     &   QND$_6$      \\
   \hline
$|\phi^{+}\rangle_{P}$   &       $\pm\theta$                      &   $\pm\theta$\\
$|\phi^{-}\rangle_{P}$   &       $\pm\theta$                      &   $0$ \\
$|\psi^{+}\rangle_{P}$   &       $0$                      &   $\pm\theta$ \\
$|\psi^{-}\rangle_{P}$   &       $0$                      &   $0$\\
\hline\hline
\end{tabular}\label{table3}
%\end{ruledtabular}
\end{table}

\section{Discussion}

In our six-qubit HBSA scheme, we exploit the cross-Kerr nonlinearity
to construct the QNDs to check the parity of the two photons in the
three DOFs. Therefore, we should acknowledge that the feasibility of
the proposed scheme depends on the nonlinear phase shift of the Kerr
media. Although many works have been reported on cross-Kerr
nonlinearity \cite{Kok,crosskerr1,crosskerr2}, a clean cross-Kerr
nonlinearity in the optical single-photon regime is quite a
controversial assumption with current technology. In 2006 and 2007,
Shapiro and Razavr \cite{crosskerr3,crosskerr4} pointed out that the
single-photon Kerr nonlinearity may do no help in quantum
computation. Moreover, in 2010, the research results of
Gea-Banacloche \cite{crosskerr5} suggested that a large phase shift
via a ``giant Kerr effect" with single-photon wave packets is
impossible at present.

Fortunately, our HBSA scheme only requires a small phase shift, as
long as it can be distinguished from zero, and much progress has
been made on the Kerr nonlinearity and homodyne detection. In 2003,
Hofmann \emph{et al.} \cite{Hofmann} demonstrated that a phase shift
of $\pi$ can be achieved with a single two-level atom one-sided
cavity system. In 2010, Wittmann \emph{et al.} \cite{Wittmann}
investigated the difference between a standard homodyne detector and
a displacement-controlled photon number resolving (PNR) detector.
They showed that the displacement-based PNR receiver outperforms the
standard homodyne detection. Therefore, for a weak cross-Kerr
nonlinearity $\theta\ll1$, if we choose a sufficiently large
amplitude of the coherent state, which satisfies the requirement
$\alpha\theta^{2}\gg1$,   it is possible for us to achieve
deterministic distinguishability between the shifted and non-shifted
phases in the coherent state. Furthermore, in 2011, He \emph{et al.}
\cite{crosskerr2} showed that effects due to the transverse degrees
of freedom significantly affect the cross-phase modulation process,
and made the treatment of single-photon-coherent-state interactions
more realistic. In the same year, Feizpour \emph{et al.}
\cite{crosskerr7} researched the cross-Kerr nonlinearity between
continuous-mode coherent states and single photons, and they
indicated that a cross-Kerr phase shift is likely to be amplified to
observable value with weak-value amplification. Moreover, Zhu and
Huang \cite{crosskerr8} showed that giant Kerr nonlinearity of the
probe and the signal pulses may be achieved with nearly vanishing
optical absorption. The substantial cross-Kerr nonlinearities
\cite{Hoi, Sathy} have already been obtained in the microwave domain
using superconducting qubits. In the work by Hoi \emph{et al.}
\cite{Hoi}, the average cross-Kerr phase shift was demonstrated up
to 20 degrees per photon with both coherent microwave fields at the
single-photon level.

Before ending this work, we will briefly discuss the application of
our HBSA scheme in quantum hyperdense coding. As quantum hyperdense
coding is the generalization of quantum dense coding with photon
systems in several DOFs, with our six-qubit HBSA scheme, one can
transfer six bits of classical information by sending only one
photon. In order to realize quantum hyperdense coding, the sender
must choose one of the local 64 operations $\{U_{I}^{P}, U_{x}^{P},
U_{y}^{P}, U_{z}^{P}\}\otimes\{U_{I}^{F}, U_{x}^{F}, U_{y}^{F},
U_{z}^{F}\}\otimes\{U_{I}^{S}, U_{x}^{S}, U_{y}^{S}, U_{z}^{S}\}$ to
perform on photon, in which $U^{i}_{j}$ ($i=P, F, S$ and $j=I, x,
y,z$) are unitary operations in polarization or one of the two
longitudinal momentum DOFs. Here, the unitary operations $U_{x}^{P}$
and $U_{z}^{P}$ can be achieved by a half-wave plate set at $45^{o}$
and $0^{o}$, respectively. The combination of $0^{o}$ and $45^{o}$
half-wave plates can be used to perform the unitary operation
$U_{y}^{P}$. One can accomplish the operation $U_{x}^{P}$,
$U_{y}^{P}$, and $U_{z}^{P}$ by putting appropriate half-wave plates
in all the four paths of the photon. $U^{i}_{I}$ ($i=P, F, S$) is
unit operation, which means doing nothing on the photon. For
single-photon longitudinal-momentum states, one can exchange the two
modes to accomplish the operation $U_{x}^{i}$ ($i=F, S$). The
operation $U_{z}^{i}$ ($i=F, S$) can be achieved by putting $0^{o}$
half-wave plates in the appropriate path. The operation $U_{y}^{i}$
($i=F, S$) is the combination of $U_{z}^{i}$ ($i=F, S$) and
$U_{x}^{i}$ ($i=F, S$). Using those operations and our six-qubit
HBSA scheme, we can accomplish the six-bit quantum hyperdense coding
which will largely improve the capacity of long-distance quantum
communication.

In summary, we have proposed an efficient scheme for the complete
nondestructive analysis of hyperentanglement of two-photon systems
in three DOFs with the help of the cross-Kerr nonlinearity. We use
cross-Kerr nonlinearity to construct quantum nondemolition detectors
which are used to make a parity-check measurement and analyze Bell
states in different DOFs of two-photon systems. We have also
presented the applications of our HBSA protocol in quantum
hyperdense coding with two-photon systems entangled in three DOFs
simultaneously, which means that our HBSA protocol  may be useful
for practical applications in quantum information processing, blind
quantum computation, distributed quantum computation, and especially
long-distance high-capacity quantum communication in future.

\section*{Methods}

\textbf{Cross-Kerr nonlinearity.}
The Hamiltonian of a cross-Kerr nonlinearity medium is \cite{Kok,Nemoto}   %  Eq 12
\begin{eqnarray}\label{eq2}
H=\hbar\chi a^{\dag}_{s}a_{s}a^{\dag}_{p}a_{p}.
\end{eqnarray}
Here $a_{s}$ ($a_{p}$) and $a^{\dag}_{s}$ ($a^{\dag}_{p}$) are the
annihilation and the creation operators of the signal (probe) pulse
beam, respectively. $\hbar\chi$ is the coupling strength of the
nonlinearity, which is decided by the property of the nonlinear
material. If we consider that the probe beam is the coherent state
$|\alpha\rangle$, for an arbitrary signal state
$|\varphi\rangle_{s}=c_{0}|0\rangle_{s}+c_{1}|1\rangle_{s}$, the
effect of the cross-Kerr nonlinearity on the whole system can be
described as
\begin{eqnarray}    %  Eq 13
\begin{split}
U|\varphi\rangle_{s}|\alpha\rangle_{p}&=e^{iH_{QND}t/\hbar}(c_{0}|0\rangle_{s}+c_{1}|1\rangle_{s})|\alpha\rangle_{p} \\
&=c_{0}|0\rangle_{s}|\alpha\rangle_{p}+c_{1}|1\rangle_{s}|\alpha
e^{i\theta} \rangle_{p},
\end{split}
\end{eqnarray}
where $|0\rangle_{s}$ and $|1\rangle_{s}$ are the Fock states for
the signal pulse. The phase shift $\theta=\chi t$ and $t$ is the
interaction time which is proportional to the number of photons with
the single-photon state being unaffected.

\section*{Acknowledgments}

FGD was supported by the National Natural Science Foundation of
China under Grant Nos. 11174039 and 11474026, and the Fundamental
Research Funds for the Central Universities under Grant No.
2015KJJCA01. QA was supported by the National Natural Science
Foundation of China under Grant No. 11505007, the Youth Scholars
Program of Beijing Normal University under Grant No. 2014NT28, and
the Open Research Fund Program of the State Key Laboratory of
Low-Dimensional Quantum Physics, Tsinghua University Grant No.
KF201502. MZ was supported by the National Natural Science
Foundation of China under Grant No. 11475021 and the National Key
Basic Research Program of China under Grant No. 2013CB922000.

\section*{Author contributions}

Q.L., G.Y., M.Z., and F.G. wrote the main manuscript text, and
prepared Figures 1-5. Q.L., Q.A., and F.G. completed the
calculations. F.G. supervised the whole project. All authors
reviewed the manuscript.

\end{document}